\documentclass[12pt]{article}
\usepackage{mathrsfs}
\usepackage{}
\usepackage{geometry}  
\usepackage{graphicx}
\usepackage{amssymb}
\usepackage{amsmath}
\usepackage{epstopdf}
\usepackage{comment}
\usepackage{cite}

\usepackage[hyperindex=true,
          pdfstartview=FitH,
          bookmarksnumbered=true,
          bookmarksopen=true,
          citecolor=blue,
          linkcolor=blue,
          colorlinks=true,
          unicode]{hyperref}

\allowdisplaybreaks

\begin{document}

\title{Three dimensional rotating hairy black holes, asymptotics and
thermodynamics}
\author{Wei Xu$^{1,2}$ \thanks{{\em
        email}: \href{mailto:xuweifuture@gmail.com}
        {xuweifuture@gmail.com}}\ ,
        Liu Zhao$^{2}$\thanks{{\em
        email}: \href{mailto:lzhao@nankai.edu.cn}{lzhao@nankai.edu.cn}}
         and De-Cheng Zou$^{3}$ \thanks{{\em
        email}: \href{mailto:zoudecheng@sjtu.edu.cn}
        {zoudecheng@sjtu.edu.cn}}\\
$^{1}$School of Physics, Huazhong University of Science and Technology, \\ Wuhan 430074, China\\
$^{2}$School of Physics, Nankai University, Tianjin 300071, China\\
$^{3}$Department of Physics and Astronomy, Shanghai Jiao Tong University, \\Shanghai 200240, China}

\date{}
\maketitle

\begin{abstract}
A rotating hairy black hole solution is found in gravity minimally coupled to a
self-interacting real scalar field in three spacetime dimensions. Then we
discuss analytically the horizon structure and find an analogue of the famous
Kerr bound in (2+1) dimensions because of the existence of black hole horizons.
We present the asymptotic symmetries and find the same symmetry group (i.e. the
conformal group) and central charge as in pure gravity. Based on the asymptotic
behavior, the mass and angular momentum are presented by the Regge-Teitelboim
approach. Other thermodynamic quantities are also obtained and the first law
of black hole thermodynamics and Smarr relation are checked. In addition, we
also investigate the local thermodynamic stability and find the existence of
Hawking-Page phase transition in the rotating hairy black hole.
\end{abstract}

\section{Introduction}

Three-dimensional gravity has been widely considered as a useful laboratory for
the understanding of gravitation in four and higher dimensions. One of the most
important examples is the BTZ black hole \cite{Banados:1992wn}, which shares
several features of higher dimensional black holes, the most interesting one
being the AdS/CFT correspondence \cite{Maldacena:1997re}. The strong evidence
for this correspondence was proposed earlier by Brown and Henneaux
\cite{Brown:1986nw}. They showed that the asymptotic conditions of
three-dimensional anti-de Sitter spaces (AdS) are left invariant under the
conformal group in two dimensions. Using this correspondence,
one can find the microscopic and semiclassical black hole entropy by means of
Cardy formula \cite{Cardy:1986ie}. For matter-free gravity theory, it was done
by Strominger for the BTZ black hole \cite{Strominger:1997eq}. A further
approach of this subject has been soon provided by Carlip
\cite{Carlip:1998qw,Carlip:1998wz,Carlip:2011vr}.
This application of AdS/CFT correspondence has been widely extended to other
different gravity theories with extra matter sources. For instance, gravity with
scalar fields \cite{Martinez:1996gn,Natsuume:1999cd,Henneaux:2002wm,
Correa:2010hf,Hasanpour:2011ji,Correa:2011dt,Correa:2012rc,Zhao:2013isa,
Xu:2013nia}, new massive gravity \cite{Bergshoeff:2009hq,Giribet:2009qz,
Perez:2011qp}, topologically massive gravity \cite{Myung:2008ey,
Henneaux:2010fy,Detournay:2012pc} and higher spin theories
\cite{Perez:2012cf,David:2012iu,Perez:2013xi}.

Recently, several hairy black holes in gravity with non-minimally coupled a
self-interacting real scalar field in three spacetime dimensions are presented
\cite{Zhao:2013isa,Xu:2013nia} and has attracted considerable attentions
\cite{Belhaj:2013ioa,Sadeghi:2013eua,Sadeghi:2013gmf,Sadeghi:2013jja,
Naji:2014ira,Xu:2014qaa,Meng:2014dfa,Wu:2014fwa},
especially for their application of holographic principle
\cite{Meng:2014dfa,Wu:2014fwa}. On the other hand, an exact hairy black hole
solution has been given in gravity with a minimally coupled self-interacting
real scalar field, in four dimensions \cite{Martinez:2004nb}.
This result is somehow unexpected since it circumvents the so called no-hair
conjecture. This makes it important to consider the similar cases in three
dimensions, i.e. the three dimensional black holes in gravity minimally coupled
to a self-interacting real scalar field. Actually, a single, minimally coupled,
scalar field is the simplest case and is very interesting.
Since there is no continuous symmetry in the Lagrangian, it is not possible to
associate it with a conserved current, and therefore is a natural candidate
for a dark matter component.

In this article we focus on some new rotating black hole of three dimensional
Einstein gravity minimally coupled to self-coupling scalar field. In this model,
the fall-off of the scalar field at infinity is slower than that of
a localized distribution of matter, while the asymptotic symmetry group remains
the same as in pure gravity (i.e., the conformal group). This means,
the generators of the asymptotic symmetries, however, acquire a contribution
from the scalar field, but the algebra of the canonical generators possesses
the standard central extension. Similar results have been found in different
static as well as rotating black holes of
this theory \cite{Henneaux:2002wm,Correa:2010hf,Correa:2011dt,Correa:2012rc,Aparicio:2012yq}.
The mass $M$ and angular momentum $J$ of the new black hole solution will be
computed using the Regge-Teitelboim approach. Furthermore,
we will consider the thermodynamics of the black hole and find
the existence of Hawking-Page phase transition for this rotating hairy black hole.

The rest of this paper is organized as follows. We first present the
new rotating hairy black hole in Section 2 and discuss its horizon structure
in Section 3. Section 4 is devoted to considering the asymptotics and computing
the conserved charges (mass and angular momentum)
using the Regge-Teiltelboim approach. In Section 5, the thermodynamics of this
rotating hairy black hole is discussed.
Finally, some concluding remarks are given in Section 6.

\section{Action, field equation and generic solution}

Consider the system with action
\begin{align}
I=\frac{1}{2}\int \mathrm{d}^{3}x\sqrt{-g}\left[
R- \nabla_{\mu} \phi\nabla^{\mu} \phi -2V(\phi )\right] \;,
\end{align}
where $\phi$ is the scalar field and $V(\phi)$ is the scalar self-coupling
potential which is given by
\begin{align}
    V(\phi)&=-\frac{1}{\ell^2}\cosh^6\left(\frac{1}{2\sqrt{2}}\phi\right)
    +\frac{1}{\ell^2}\left(1+\mu\ell^2\right)
    \sinh^6\left(\frac{1}{2\sqrt{2}}\phi\right)\nonumber\\
    &-\frac{\alpha^2}{64}\sinh^{10}\left(\frac{1}{2\sqrt{2}}\phi\right)
    \cosh^{6}\left(\frac{1}
    {2\sqrt{2}}\phi\right)\bigg[\tanh^6\left(\frac{1}{2\sqrt{2}}\phi\right)
    \nonumber\\
    &-5\tanh^4\left(\frac{1}{2\sqrt{2}}\phi\right)
    +10\tanh^2\left(\frac{1}{2\sqrt{2}}\phi\right)-9\bigg],
\end{align}
where $\ell$, $\mu$ and $\alpha$ are all constants.

\begin{figure}[h]
\centering
\begin{minipage}{.45\textwidth}
\centering
\includegraphics[width=\linewidth]{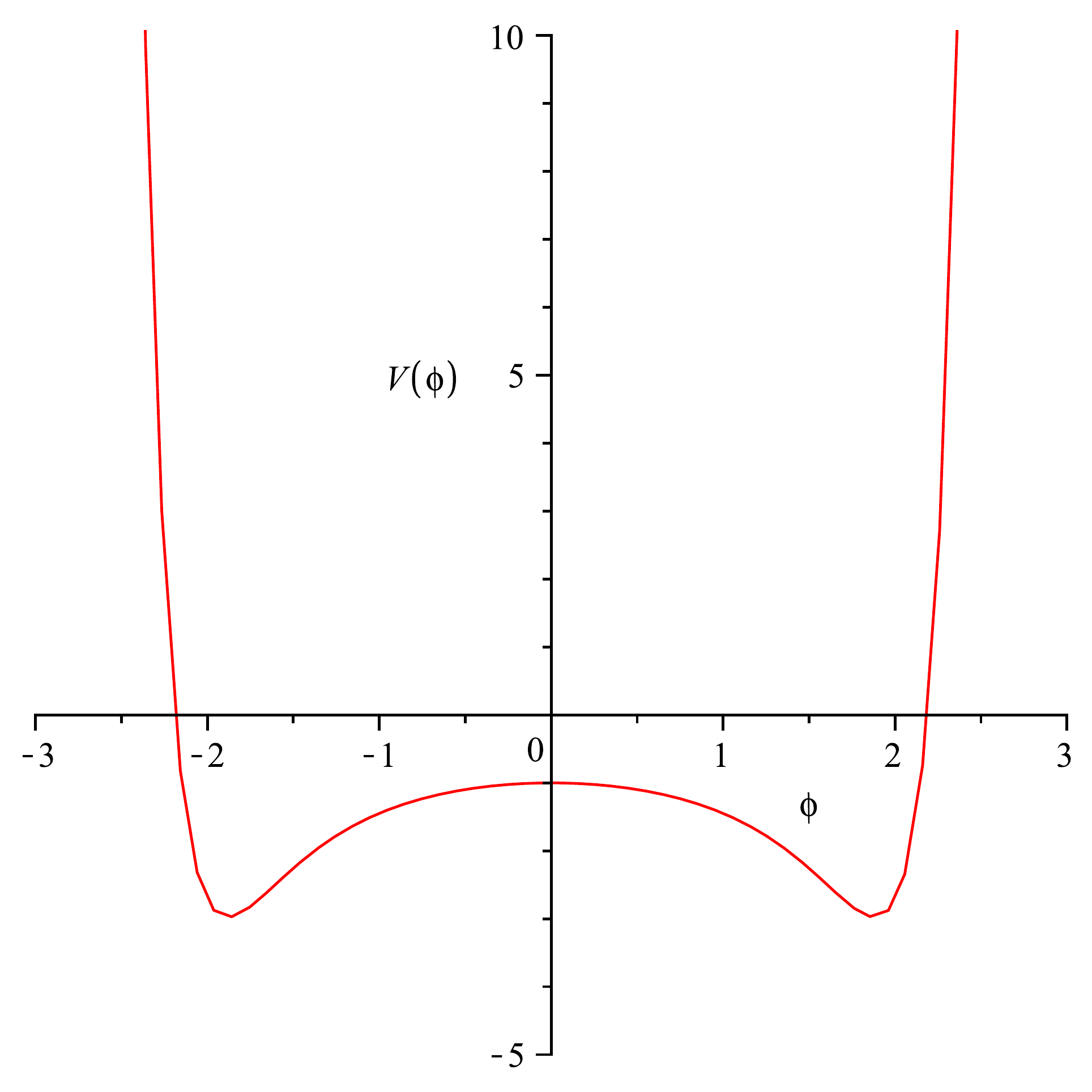}
\caption{Plot of scalar potential $V(\phi)$ versus $\phi$, with $\ell=1$, $\mu=-3$, $\alpha=1$.}
\label{FigV(phi)}
\end{minipage}\hspace{15pt}
\end{figure}

To have an intuitive feeling about the property of the scalar potential, we
create a plot of $V$ versus $\phi$ at $\ell=1$, $\mu=-3$ and $\alpha=1$ as
presented in Fig.\ref{FigV(phi)}. Meanwhile, we can also make a power series
expansion for the scalar potential, which turns out to read
\begin{align}
  V(\phi)=&-\frac{1}{\ell^2}-\frac{3}{8\ell^2}\phi^2
  -\frac{1}{16\ell^2}\phi^4
  -\frac{32-15\mu\ell^2}{7680\ell^2}\phi^6
  -\frac{64-105\mu\ell^2}{430080\ell^2}\phi^8\nonumber\\
  &-\frac{512-2205\mu\ell^2}{154828800\ell^2}\phi^{10}]
  +\frac{9\alpha^2}{2097152}\phi^{10}+O(\phi^{12}).
\end{align}
The zeroth order term plays the role of a negative cosmological constant
$\Lambda=-\frac{1}{\ell^2}$, which is required for the existence of a smooth
black hole horizon in three dmensions \cite{Ida:2000jh}. The second order term
is a mass term with negative mass squared $m^2=-\frac{3}{4\ell^2}$, which
obeys the Breitenlohner-Freedman bound in three dimensions \cite{P1982,M1985}.
From the tenth order and onwards, the parameter $\alpha$ begins to make
contributions, which ensures that the potential is bounded from below and
possesses three extrema: one local maximum at $\phi=0$
with $V(0)=\Lambda$ and two minima at $\phi=\pm\phi_0$ for some $\phi_0$,
with $V(\pm\phi_0)<\Lambda$ (however, it is difficult to obtain $\phi_0$
analytically).

The equations of motion (EOM) read as
\begin{align}
  &G_{\mu\nu}-\left(\partial_{\mu}\phi\partial_{\nu}\phi
  -\frac{1}{2}g_{\mu\nu}\nabla^{\rho}\phi\nabla_{\rho}\phi\right)
  +V(\phi)g_{\mu\nu}=0,\label{Einstein}\\
  &g^{\mu\nu}\nabla_{\mu}\nabla_{\nu}\phi
  -\partial_{\phi}V(\phi)=0.\label{scalar}
\end{align}
We choose the rotating metric anstaz
\begin{align}
\mathrm{d}s^{2}=-\mathcal {A}\mathrm{d}t^{2}
+\frac{\mathrm{d}\rho^{2}}{\mathcal {B}}
+\rho^{2}\bigg(\mathrm{d}\psi-\omega\mathrm{d}t\bigg) ^{2},
\end{align}
where $\mathcal {A}, \mathcal {B}$ and $\omega$ are all functions of the radial
coordinate $\rho$ only. Plugging the metric into the EOM, we can obtain the
scalar field
\begin{align}
\phi(\rho) =2\sqrt{2}{\rm arctanh}\sqrt{\frac{B}{H(\rho)+B}},\quad
H(\rho)=\frac{1}{2}\left( \rho+\sqrt{\rho^{2}+4B\rho}\right)\label{phisol}
\end{align}
and the functions $\mathcal {A}$ and $\mathcal {B}$ appearing in the metric
ansatz:
\begin{align}
&\mathcal {A}=\left( \frac{H}{H+B}\right) ^{2}f(H),\quad
\mathcal {B}=f(H)\left( \frac{H+2B}{H+B}\right) ^{2},\\
&f(H)=3\mu B^2+\frac{2\mu B^3}{H}+\frac{\alpha^2 B^4(3H+2B)^2}{H^4}
+\frac{H^2}{\ell^2},\label{fsol}\\
&\omega(H)=\frac{\alpha B^2(3H+2B)}{H^3}. \label{omegasol}
\end{align}
Here the solution is characterized by four parameters: $\ell,\mu,B,\alpha$.
The parameter $B$ needs to be non-negative in order to keep the scalar field
real. Meanwhile, we need to take $\mu<0$ in order to have a positive
mass for the black hole solution, as will be shown later.
When $B$ or $\alpha$ is vanishing, we can find two well-known degenerate cases:
when $B=0$, the solution degenerates to the massless static BTZ solution
\cite{Banados:1992wn}; by setting $\mu\ell^2=-\nu-1$,
the solution with $\alpha=0$ degenerates to the static hairy black
hole considered in \cite{Henneaux:2002wm,Correa:2010hf,Correa:2011dt,
Nadalini:2007qi}. After a conformal transition, the full solution can be turned
into the one found in the non-minimal coupling scalar-tensor theory of gravity
\cite{Zhao:2013isa}, as is shown in the Appendix. Note that applying a conformal transformation to the solutions in the minimally
coupled theory, the solution obtained in the Jordan frame are not always physically
equivalent to the untransformed ones \cite{Faraoni:1998qx}.

To understand the geometric characteristics of
the solution, we calculate the Ricci scalar of the solution. The result reads
\begin{align}
  R=&-\frac{6}{\ell^2}-\frac{16B}{H^2\ell^2}(B+H)
  +\mu B^3\left( \frac {4{B}^{2}}{{H}^{5}}+\frac {10{B}}{{H}^{4}}+
\frac {12}{{H}^{3}}\right) \nonumber\\
&+ 2\alpha^2B^5\left( \frac {13{B}^{3}}{{H}^{8}}
+\frac {54{B}^{2}}{{H}^{7}}+\frac {72{B}}{{H}^{6}}
+\frac {36}{{H}^{5}} \right).
\end{align}
It is clear that the solution has an essential singularity at $H = 0$
whenever $B\neq0$. Higher order curvature invariants
such as $R_{\mu\nu}R^{\mu\nu}$ and $R_{\mu\nu\lambda\sigma}
R^{\mu\nu\lambda\sigma}$ have the same behavior. Note that for $\rho\ge 0$,
$H(\rho)=0$ implies $\rho=0$. On the other hand, we find that some of the
components of the Cotton tensor, e.g.
\begin{align}
  C_{\psi\rho\psi}=3B^3\left(\frac {\mu}{{H}^{2}}
  +\frac {6B^2{\alpha}^{2}}{{H}^{4}}
  +\frac {4{B}^{3}{\alpha}^{2}}{{H}^{5}}\right)
\end{align}
are nonvanishing if $B\neq0$, signifying that the metric is
non conformally flat \cite{Garcia}.

\section{The horizon structure}

In this section, we present the horizon structure analytically.
The horizons must be situated at the zeros of the function $f(H)$.
We are mostly interested in the event horizon of the black hole,
which corresponds to the largest zero $H_+=H(\rho=\rho_+)$ of $f(H)$.
Denoting
\begin{align}
  X_0=\frac{3H_{+}+2B}{H_+^3}, \label{X_0}
\end{align}
which never vanishes for $H_+>0$, we can rearrange $f(H_+)$ in the form
\begin{align}
  f(H_+)=\frac {{H_+}^{2}}{{\ell}^{2}} \left( 1+\mu B^2\,{\ell}^{2}X_0
  +{\alpha}^{2}{\ell}^{2}B^4{X_0}^{2}
 \right) =0.
\end{align}
We can directly get two roots:
\begin{align}
&X_0^{(1)}=-\,{\frac {\mu\ell+\sqrt {{\mu}^{2}{\ell}^{2}
  -4\,{\alpha}^{2}}}{2{\alpha}^{2}\ell B^2}}=\frac{\tilde{X}_0^{(1)}}{B^2},\\
&X_0^{(2)}=-\,{\frac {\mu\ell-\sqrt {{\mu}^{2}{\ell}^{2}
-4\,{\alpha}^{2}}}{2{\alpha}^{2}\ell B^2}}=\frac{\tilde{X}_0^{(2)}}{B^2}.
\end{align}
Note that only when $\mu\leq-\frac{2\alpha}{\ell},$ $X_0^{{i}} (i=1,2)$ are
real-valued. If $-\frac{2\alpha}{\ell}<\mu<0$, $X_0^{{i}}$ $(i=1,2)$ are both
imaginary.

Now consider $\tilde{X}_0^{(i)} (i=1,2)$ as a constant, we can rewrite
Eq.(\ref{X_0}) as
\begin{align}
  g(h)\equiv \tilde{X}_0^{(i)}h^3-3h-2=0, \label{HH}
\end{align}
where we have set
\begin{align}
H_+=h B .\label{h}
\end{align}
One can analytically solve Eq.\eqref{HH} and get three solutions,
\begin{align}
  &h^{(1)}=\frac{X_1}{\tilde{X}_0^{(i)}}+\frac{1}{X_1},\nonumber\\
  &h^{(2,3)}=-\frac{1}{2}\bigg(\frac{X_1}{\tilde{X}_0^{(i)}}
  +\frac{1}{X_1}\bigg)
  \pm I\frac{\sqrt{3}}{2}\bigg(\frac{X_1}{\tilde{X}_0^{(i)}}
  -\frac{1}{X_1}\bigg),
\label{H_+}
\end{align}
where
\begin{align*}
  X_1= \left[\left(1+\sqrt{\frac {-1+\tilde{X}_0^{(i)}}{\tilde{X}_0^{(i)}}}
  \right)\left(\tilde{X}_0^{(i)}\right)^{2}\right]^{1/3}.
\end{align*}

Consider function $g(h)$ as given in \eqref{HH} with
$\tilde{X}_0^{(i)}>0$. It is evident that  $g(0)<0,g(+\infty)>0$,
and for positive $h$, $g'(h)$ starts with the negative value $-3$ at $h=0$ and
increases monotonically until it becomes positive at some $h$. Therefore,
near $h=0$, $g(h)$ firstly decreases from a negative value to the minimum,
and then increases monotonically. Hence the curve $g(h)$ will only come across
the horizontal axis only once, which implies that only a single solution
of Eq.\eqref{HH} is real and positive for fixed positive $\tilde{X}_0^{(i)}$.
Finally, by inverting the $H(\rho)$ relation given in \eqref{phisol}, we can
express the black hole horizon radius as
\begin{align}
  \rho_+=\frac{H_+^2}{H_++B}=\frac{h^2}{h+1}B,
\label{rho_+}
\end{align}
where $h$ is to be taken as the real positive solution to \eqref{HH}. It is
clear in this parametrization of the horizon radius that the horizon radius
is proportional to the parameter $B$ because $h$ is independent of $B$.

Recall that, depending on a bound over the parameters $\mu, \ell$ and $\alpha$,
there can be three different cases for the values of $\tilde{X}_0^{(i)}$.
These different cases will also affect the number of real positive roots for
$g(h)$:
\begin{enumerate}
\item $\mu<-\frac{2\alpha}{\ell}$: in this case,
$\tilde{X}_0^{(i)} (i=1,2)$ are both positive and real, each results in
a real positive $h$. Among these, the horizon corresponding to the choice
$\tilde{X}_0^{(1)}$ is the outer horizon.
This case corresponds to the non-extremal rotating hairy black hole;
\item $\mu=-\frac{2\alpha}{\ell}$: in this case, $\tilde{X}_0^{(i)} (i=1,2)$
coincides and is real positive, which results in a single positive real $h$.
This case corresponds to the extremal rotating hairy black hole;
\item $-\frac{2\alpha}{\ell}<\mu<0$:
in this case, $\tilde{X}_0^{{i}} (i=1,2)$ are both imaginary,
which can never result in a real $h$ from Eq.(\ref{HH}).
This case corresponds to a naked singularity.
\end{enumerate}

Summarizing the above cases, we find that the existence of black hole
horizons requires a bound over the parameters, i.e.
\begin{align}
  -\frac{\mu}{\alpha}\geq\frac{2}{\ell}.
\label{bound}
\end{align}
This bound will be shown to be an analogue of the famous Kerr bound for rotating
black holes in the next section.

\section{Asymptotics and  conserved charges}

The asymptotic behavior of our solution belongs to the following class:
\begin{align}
&\phi =2\sqrt{2}\left(\frac{B^{1/2} }{\rho^{1/2}}
-\frac{2}{3}\frac{B^{3/2}}{\rho^{3/2}}\right)
+\mathcal {O}(\rho^{-5/2}),\label{AsymptPhi}
\\
&g_{tt}= -\frac{\rho^{2}}{\ell^{2}}+\mathcal {O}(1), &&
g_{\rho\rho} = \frac{\ell^{2}}{\rho^{2}}
-\frac{4\ell^{2}B}{\rho^{3}}+\mathcal {O}(\rho^{-4}),  \\
& g_{t\rho} = \mathcal {O}(\rho^{-2}),
&& g_{\psi\psi } =  \rho^{2}+O(1), \\
& g_{\psi \rho} = \mathcal {O}(\rho^{-2}),
&& g_{t\psi } =\frac{3\alpha B^2}{\rho^2}-\frac{4\alpha B^3}{\rho^3}
+\mathcal {O}(\rho^{-4}).
\label{AsymptGeneral}
\end{align}
Although $g_{t\rho }$, $g_{\psi\rho }$ and $g_{\rho\rho}$ falls off
slower than in pure gravity, notice that this set of conditions is shown to be left
invariant under the Virasoro algebra generated by the asymptotic Killing vectors of three dimensional pure AdS gravity \cite{Henneaux:2002wm}.

Using the Regge-Teitelboim approach \cite{Regge:1974zd},
the contributions to the conserved charges of pure gravity and the scalar field: $
Q_{G}(\xi )$ and $Q_{\phi }(\xi )$, can be found.
With the help of the asymptotic conditions Eq.(\ref{AsymptGeneral}),
their variations can be shown to
be of the form \cite{Henneaux:2002wm}
\begin{align}
\delta Q_{G}(\xi ) &=\frac{1}{2 }\int d\psi \left\{ \xi ^{\bot }\left[ \frac{2}{\ell\rho}\delta
g_{\psi \psi }+\rho g^{-1/2}g^{\rho\rho}\delta g_{\rho\rho}\right]  +2\xi
^{\psi }\delta g_{\psi t}\right\}  \nonumber \\
&=\frac{1}{2 }\int d\psi \left\{
\frac{1}{\ell\rho}\xi ^{\bot }\delta g_{\psi \psi }-2\rho\xi ^{\bot }\delta (g^{-1/2})
+2\xi ^{\psi }\delta g_{\psi t}\right\}\label{DeltaQG} \;.
\\
\delta Q_{\phi }(\xi ) &=-\int d\psi \bigg\{ \xi ^{\bot }g^{1/2}g^{\rho\rho}\partial
_{\rho}\phi \delta \phi \bigg\} \label{DeltaQPhi} \;.
\end{align}
Asymptotically, $\xi ^{\bot }\sim \xi ^{t}\rho/\ell$. One can find that
the generators of the asymptotic symmetries has a non-vanishing contribution from the scalar field.

Integrating to obtain the total charge $Q=Q_{G}+Q_{\phi }$, it reads \cite{Henneaux:2002wm}
\begin{align}
Q(\xi )=\int d\psi\left\{ \frac{1}{2}\left[ \frac{\xi ^{\bot }}{\rho\ell}\left(
(g_{\psi \psi }-\rho^{2})-2\rho^{2}(\ell g^{-1/2}-1)  \right)
+2\xi ^{\psi }\pi _{\psi }^{\rho}\right] +4\rho^{2}\left[ \phi ^{2}+
\frac{1}{3}\phi ^{4}\right]\right\},
\label{Qtotal}
\end{align}

Note that one can find the algebra of the charges (\ref{Qtotal}) is identical to the one
found in \cite{Brown:1986nw}, namely, two copies of the Virasoro algebra with a central
extension. Then computing the variation of the charges on the vacuum,
one can find the central charge. Here the vacuum is chosen to be the same as in pure
gravity, which shows that the value of the central charge remains unchanged
\begin{align}
c=\frac{3\ell}{2G}\,.  \label{Central Charge}
\end{align}
Based on this central charge, one can easily obtain the semiclassical area entropy of the
rotating hairy black hole after applying Cardy formula.

Using Eq.(\ref{Qtotal}),
the mass $M$ and the angular momentum $J$ of the hairy black hole described
by Eqs.(\ref{fsol},\ref{omegasol}) are found to be
\begin{align}
&M\equiv Q(\partial _{t})=-3\mu\,B^2, \label{M} \\
&J\equiv Q(\partial _{\psi})={6\alpha}B^2. \label{a}
\end{align}
Inserting them into Eq.(\ref{bound}), one will find the real Kerr-like bound in (2+1)
dimensions,
\begin{align}
  \frac{M}{J}\geq\frac{1}{\ell}.
\end{align}

\section{Thermodynamics of rotating hairy black hole}

In this section, we consider the thermodynamics and phase transition of this rotating hairy black hole
in minimally coupling model. From Eqs.~(\ref{M})(\ref{a}), the rotating hairy black hole solution
can be rewrite to be
\begin{align*}
  f(H)=-M-\frac{2BM}{3H}+\frac{J^2 (3H+2B)^2}{36H^4}
+\frac{H^2}{\ell^2}.
\end{align*}
Considering Eq.(\ref{h}), the mass of rotating hairy black hole can be obtained as
\begin{align}
  M=\,{\frac {{J}^{2}{\ell}^{2}(3h+2)^2+36h^2\,H_{+}^{4}}{12{\ell}^{2} \left( 3h+2\right)H_{+}^{2} }}.
\end{align}

The Hawking temperature of the rotating hairy black hole can be obtained by means of the surface gravity
$\tilde{\kappa}^2=-\frac{1}{2}(\nabla^a \chi_b )(\nabla_a \chi^b )$ \cite{Carroll:2004st},
where the Killing vector $\chi^a $
equals $(1,0,\Omega)$ and
\begin{align}
\Omega \equiv \frac{\mathrm{d}\psi}{\mathrm{d}t}\bigg|_{\rho=\rho_+}=\omega(H_+)=\frac{(3h+2)J}{6h H_+^3}
\end{align}
is the angular velocity at the horizon. Then, the Hawking temperature $\kappa/(2\pi)$ reads
\begin{align}
T=\,{\frac {3 \left(h+1 \right)H_{+} }{2\pi{\ell}^{2} \left( 3\,
h+2 \right)  }}-\,{\frac { \left(h+1 \right)  \left( 3h+2 \right) {J}^{2}}{24
\pi h^2\,H_{+}^{3}}}. \label{temp}
\end{align}
In addition, the Bekenstein-Hawking entropy is defined in terms of the horizon area
\begin{align}\label{BHentro}
S&=4\pi\rho_{+}=\frac{4\pi h}{(h+1)}H_+.
\end{align}
After a little calculation, one can find $T=\left(\frac{\partial M}{\partial S}\right)_{J}, \Omega=\left(\frac{\partial M}{\partial J}\right)_{S}$. Hence, it is easy to verify that the first law of black hole thermodynamics,
\begin{align}
dM=TdS+\Omega \, dJ \, ,
\end{align}
holds. Meanwhile, we can obtain the
generalized three dimensional Smarr relation for the black hole
 \begin{align}
M-\Omega J=\frac{1}{2}TS,
\label{Smarr}
\end{align}
which is the same with the one get from the scaling argument \cite{Banados:2005hm}.

\begin{figure}[h!]
\begin{center}
\includegraphics{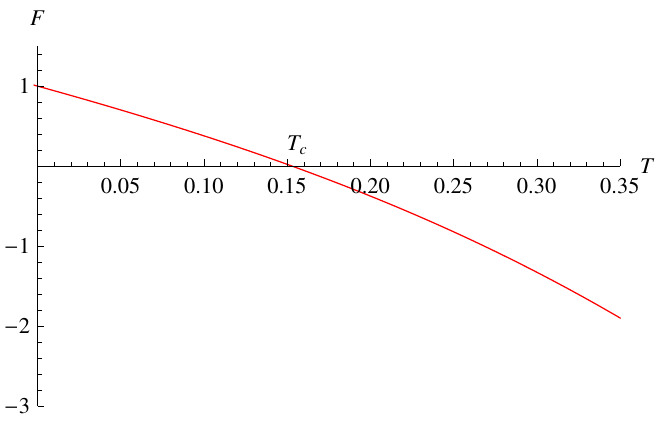}
\caption{The free energies $F$ of rotating hairy black hole versus the temperature
$T$ with $h=1$, $J=1$ and $\ell=1$.}
\label{F-T}
\end{center}
\end{figure}

In general, the local thermodynamic stability is determined by the specific heat, which states that
the positive specific heat can guarantee a stable black hole to exist, while the negative one
will make the black hole disappear when a small perturbation is included within. Here the specific
heat $C_J=T\left(\frac{\partial S}{\partial T}\right)_J$ can be expressed as
\begin{align}
  C_J=\,{\frac {96{\pi }^{2}{\ell}^{2}h^3 \left(h+2 \right)  \left( 3h
  +2 \right) ^{2}H_{+}^{4}T}{\xi\left(h+1 \right) ^{2}}},\label{C}
\end{align}
where $\xi={J}^{2}{\ell}^{2} \left( 10+20h+9h
^{2} \right)  \left( 3h+2 \right) ^{2}+36h^2 \left(
3h^{2}+4h+2 \right)\,H_{+}^{4}>0$.
As $T\geq0$, one will always obtain a positive specific heat. Therefore, the rotating hairy
black hole is always locally thermal stable in the region of $H_+>H_{ext}(\rho_{+}>\rho_{ext})$.
On the other hand, the behavior of free energy $F$ is important to determine the
phase transition of the charged hairy black hole. The free energy $F=M-TS$ reads as
\begin{align}
F=\,{\frac {\,{J}^{2}{\ell}^{2}(3h+2)^2-12h^2\,H_{+}^{4}}{4{\ell}^{2} \left( 3h+2 \right)H_{+}^{2} }}.\label{F}
\end{align}
Its zero is located in
\begin{align}
  H_+^{c}=\frac{1}{{12}^{1/4}}\,\sqrt {{\frac {J\ell \left( 3\,h+2 \right)}{h}}},
\end{align}
and the critical temperature is
\begin{align}
  T_c=\frac{\, \left( h+1 \right)}{{12}^{1/4}\pi{\ell}^{3/2}} \sqrt {{\frac {J}{ \left( 3
\,h+2 \right) h}}}.
  \label{T_c}
\end{align}
One can see that for the rotating hairy black hole,
the free energy changes its sign at at the temperature $T=T_c$, which implies the existence of
the Hawking-Page phase transition at $T=T_c$ for the rotating hairy black hole.
Moreover, the rotating hairy black holes only exist in the case of temperature $T<T_c$,
while what remain is nothing but the thermal radiation in pure AdS space when $T>T_c$. To have a clear understanding, the $F-T$ diagram with $h=1$, $J=1$ and $\ell=1$ is displayed in the Fig.\ref{F-T}. For this case, Eq.(\ref{T_c}) gives $T_c\simeq0.1529$.

\section{Conclusion}

In this paper, we present a new rotating black hole of three dimensional Einstein gravity
with minimally coupled self-coupling scalar field. Then we
discuss analytically the horizon structure and obtain an analogue of the famous
Kerr bound in (2+1) dimensions because of the existence of black hole horizons.
In this case, we also found that
the fall-off of the scalar field at infinity is slower than that of
a localized distribution of matter, while the asymptotic symmetry group remains
the same as in pure gravity (i.e., the conformal group). This means,
the generators of the asymptotic symmetries, however, acquire a contribution
from the scalar field, but the algebra of the canonical generators possesses
the standard central extension.
Furthermore, we calculate the conserved charges: mass $M$ and angular momentum $J$
by the Regge-Teitelboim approaches.

Later, we further discussed the thermodynamical properties and present
other thermodynamic quantities, such as
temperature, entropy, specific heat and free energy of the rotating black hole.
The first law of black hole thermodynamics and Smarr relation are also checked.
Then we found the rotating hairy black hole in minimally
coupling model is always locally thermal stable in the region of $\rho_+>\rho_{ext}$.
Moreover, there exist the Hawking-Page phase transition for this rotating hairy black hole
at the temperature $T=T_c$.

Note the solution we presented in this paper is
a rotating uncharged hairy black hole. However, the rotating charged hairy black
holes are difficult to obtain, while only a black hole solution for
infinitesimal electric charge and rotation parameters is given in non-minimal coupling model \cite{Sadeghi:2013gmf}.
These black hole solutions (including ones in other gravity models with matter source)
are expected to play some roles in the further studying of AdS/CFT duality.
These are all left as a future task.

\section*{Appendix: turning to non-minimal coupling model}
Based on the rotating hairy black hole in minimal coupling model, we can get the solution
in non-minimal coupling model through the following transformation
\begin{align}
&\hat{g}_{\mu \nu }=\Omega ^{2}g_{\mu \nu },\label{tranformation1}\\
&\Omega =\cosh^2(\sqrt{8}\phi),\label{tranformation2}\\
&\hat{\phi}=\sqrt{8}\tanh(\sqrt{8}\phi),\label{tranformation3}\\
&U(\hat{\phi})= V(\phi)\Omega^{-3}.\label{tranformation4}
\end{align}
where $\hat{\phi}$ is the new scalar field and $U$ is the scalar potential in the non-minimally
coupling theory. They are given by Eq.(\ref{tranformation3}) and Eq.(\ref{tranformation4}),
respectively. For other two equations, Eq.(\ref{tranformation1}) shows the conformal transformation
between the metric in two kinds of theory, Eq.(\ref{tranformation2}) is the expression
of the conformal transformation.

Then we turn to the non-minimally coupling model with the action
\begin{align*}
\hat{I}=\frac{1}{16\pi G}\int\mathrm{d} x^3\sqrt{-\hat{g}}\left[
  \hat{R}+\frac{2}{\ell^2} -\hat{g}^{\mu\nu}\nabla_{\mu}\hat{\phi}\nabla_{\nu}\hat{\phi}
  -\frac{1}{8} \hat{R}\hat{\phi}^2-2\,U(\hat{\phi})\right],
\end{align*}
The scalar potential in the action is given by \cite{Zhao:2013isa}
\begin{align}
  U(\hat{\phi})&=X \hat{\phi}^{6}+Y\,\frac {\left({\hat{\phi}}^{6}-40\,{\hat{\phi}}^{4}
  +640\,{\hat{\phi}}^{2}
  -4608\right){\hat{\phi}}^{10}}{\left({\hat{\phi}}^{2}-8\right)^{5}},
\end{align}
where
\begin{align*}
&X=\frac{1}{512}\left(\frac{1}{\ell^2}+\frac{\beta}{B^2}\right),\qquad
Y=\frac{1}{512}\left(\frac{a^2}{B^4}\right).
\end{align*}
with $\beta=\mu B^2$ and  $a=\alpha B^2$.
The rotating solution in the non-minimally coupling theory is known recently in \cite{Zhao:2013isa},
with the line element
\begin{align}
  \mathrm{d} s^2=-\hat{f}(r)\mathrm{d} t^2+\frac{1}{\hat{f}(r)}
  \mathrm{d} r^2+r^2\bigg(\mathrm{d} \psi-\hat{\omega}(r)\mathrm{d} t\bigg)^2,
 \end{align}
and all the functions in metric are
\begin{align*}
    \hat{\phi}(r)&=\pm \left(\frac{8B}{(r+B)}\right)^{1/2}, \\
    \hat{f}(r)&=3\beta+\frac{2B\beta}{r}+\frac{(3r+2B)^2a^2}{r^4}+\frac{r^2}{\ell^2},\\
   \hat{\omega}(r)&=\frac{(3r+2B)a}{r^3}.
\end{align*}

\providecommand{\href}[2]{#2}\begingroup
\footnotesize\itemsep=0pt
\providecommand{\eprint}[2][]{\href{http://arxiv.org/abs/#2}{arXiv:#2}}
\endgroup

\end{document}